\documentclass[10pt]{article}

\usepackage{geometry}
\usepackage{cite}
\usepackage{pdfcomment}
\usepackage{graphicx}
\usepackage[usenames, dvipsnames]{color}
\usepackage{soul}
\usepackage{multicol}
\usepackage{sectsty}
\usepackage{wrapfig}
\usepackage{authblk}
\usepackage[skins]{tcolorbox}
\usepackage{graphicx, type1cm, lettrine}

\sectionfont{\large}
\subsectionfont{\normalsize}

\title{\textbf{Accelerator Virtualization in Fog Computing:}\\Moving From the Cloud to the Edge}

\author[1]{Blesson Varghese\thanks{Corresponding Author; Email: b.varghese@qub.ac.uk; Web:  \url{http://www.blessonv.com}}}
\affil[1]{\normalsize School of Electronics, Electrical Engineering and Computer Science\\Queen's University Belfast, UK}
\author[1]{Carlos Rea\~no}
\author[2]{Federico Silla}
\affil[2]{Department of Computer Engineering, Universitat Politecnica de Valencia, Spain}

\date{}

\begin{document}
\maketitle

\begin{multicols}{2}
[
\vspace{-24pt}
\renewcommand{\abstractname}{\vspace{-\baselineskip}}
\begin{abstract}

\vspace{-12pt}
\begin{tcolorbox}[
    width=0.85\textwidth,
    enhanced,
    drop fuzzy shadow southwest,
    colframe=blue!50!black,
    colback=blue!2,]
Hardware accelerators are available on the Cloud for enhanced analytics. Next generation Clouds aim to bring enhanced analytics using accelerators closer to user devices at the edge of the network for improving Quality-of-Service by minimizing end-to-end latencies and response times. \textcolor{black}{The collective computing model that utilizes resources at the Cloud-Edge continuum in a multi-tier hierarchy comprising the Cloud, the Edge and user devices is referred to as Fog computing}. This article identifies challenges and opportunities in making accelerators accessible at the Edge. A holistic view of the Fog architecture is key to pursuing meaningful research in this area.
\end{tcolorbox} 

\end{abstract}

]

\thispagestyle{plain}
\pagestyle{plain}

\section*{Introduction}
\label{sec:introduction}
The Internet paved the way for providing computing as a utility service to customers. Customers pay for Cloud-based services ranging from infrastructure to software services. Infrastructure services are offered via Virtual Machines (VMs), a virtualization technology that makes remote servers accessible to users. Software services make specific applications, such as \textcolor{black}{Augmented Reality (AR)} available to users. 

\begin{figure*}[t]
  \includegraphics[width=\textwidth]{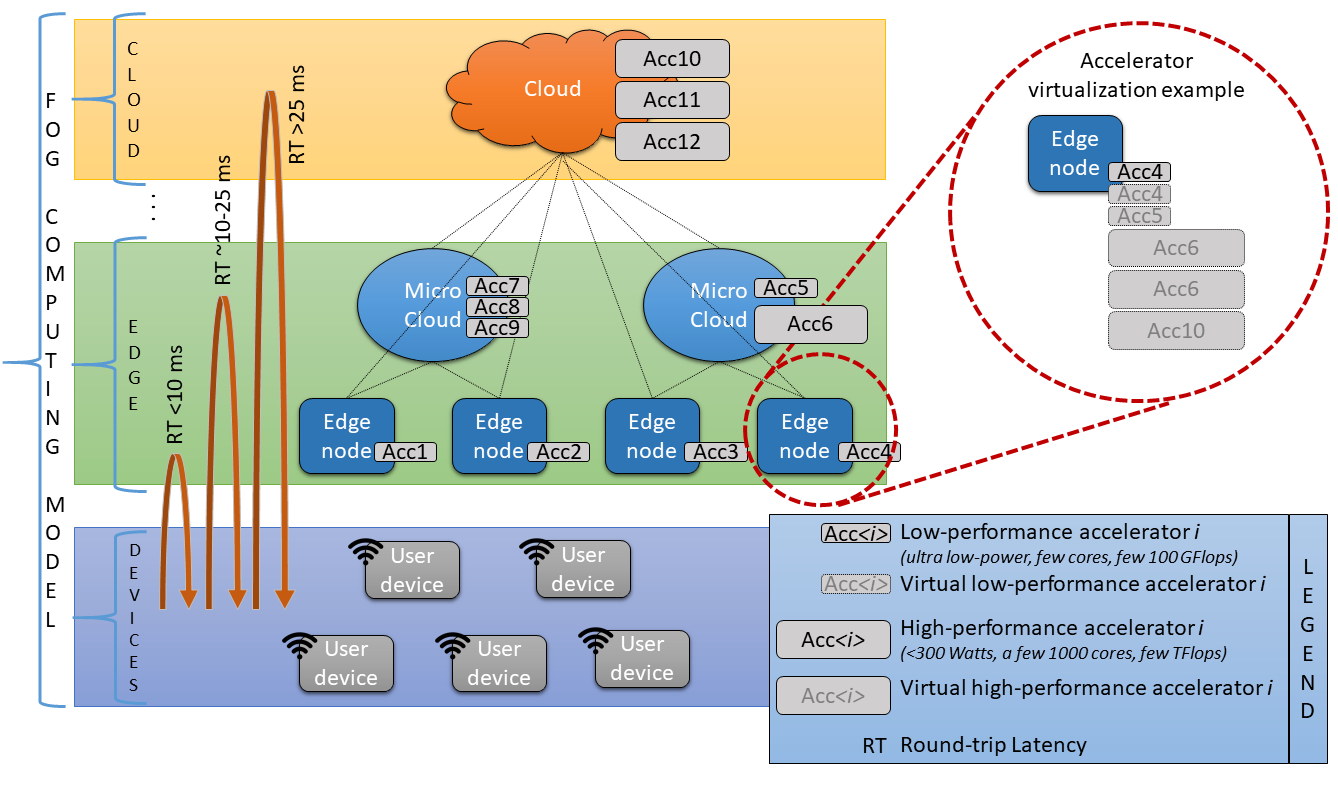}
  \caption{\textcolor{black}{A multi-tiered Fog computing model using accelerators}. The user device layer at the bottom, the Edge layer comprising nodes and micro-cloud in the middle, and the Cloud layer at the top. The Edge nodes host weak accelerators, \textcolor{black}{which may be located along with routers or wireless access points}. Edge micro-clouds that provide support for larger accelerators \textcolor{black}{ may be hosted on base stations or facilities of organizations}.Support for accelerator virtualization is highlighted for the Edge node hosting `Accelerator 4'; the Edge node can access the physical accelerator hosted on it as well as make use of a single or multiple instances of the virtual accelerator. Remote virtual accelerators located on the micro-cloud or the Cloud can also be used.} 
  \label{fig:fog}
\end{figure*}

Many services benefit from the use of accelerators, such as Graphics Processing Units (GPUs), which are specialized processors that reduce execution times of workloads on the Cloud. GPUs offer significant computing power for a relatively small power per FLOP (Floating-Point Operation) ratio.

Computational resources offered by traditional Clouds are hosted in centralized facilities, referred to as data centers. At present, traffic in the form of images, videos and text generated by user devices is usually transmitted to the Cloud for processing.
\textcolor{black}{For example, consider an AR use-case for cognitive wearable assistance in which a smart glass worn by a person provides instructions for navigating. Since the environment rapidly changes there is always a steady stream of images that needs to be processed in the Cloud.}
This will simply not be sustainable in the future. 

\vspace{2pt}
\begin{tcolorbox}[
    width=0.49\textwidth,
    enhanced,
    drop fuzzy shadow southwest,
    colframe=blue!50!black,
    colback=blue!2,]
It is anticipated that tens of billions of user devices, such as smartphones, tablets, appliances and wearables will be connected to the Internet in the next decade and consequently trillions of gigabytes of data will be generated~\cite{edgecomputing-01}. It would be practically impossible to transmit the traffic generated by all devices to the network core for processing. 
\end{tcolorbox} 

\textcolor{black}{Using the Cloud for the AR use-case will not meet real-time requirements due to latencies for rapidly responding to the change in environment as captured by the smart glass. For achieving latencies below 1 millisecond, and given the speed of light, servers must be located within 93~miles from every device in the best case.} Therefore, a fundamentally new design of the Cloud is required. 

Redesigning data centers is a central challenge in achieving sustainability for next generation Clouds~\cite{edgecomputing-02}. One strategy is to decentralize resources, so that computing can be made available at the edge of the network, which is closer to users than a distant data center. Data generated by user devices can be processed at the Edge on existing nodes, such as routers, switches or wireless access points by adding low power and low cost computing capabilities on them or by using a small dedicated cluster of compute nodes in the form of micro-clouds. Processing user requests at the Edge will reduce end-to-end latency, reduce traffic beyond the first hop of the network and improve overall Quality-of-Service (QoS)~\cite{edgecomputing-03}. 
\textcolor{black}{Instead of using the Cloud for the AR use-case Edge resources can be used to reduce latencies and minimize traffic towards the Cloud}.
Furthermore, an Edge-based approach will pave the way for a more sustainable Internet.

\textcolor{black}{Utilizing compute resources at the Cloud to Edge continuum in a multi-tiered hierarchy comprising the Cloud, the Edge and user devices is referred to as Fog computing~\cite{fogcomp-1}}. The objective of Fog computing is not to replace the Cloud, but rather to make resources, including accelerators, available closer to user devices for enhanced analytics. Figure~\ref{fig:fog} illustrates a Fog computing model using accelerators. There are three distinct layers. The bottom layer comprises user devices. The hierarchical Edge layer comprises Edge nodes that host accelerators, which are weak both in terms of power and computational capabilities, and micro-clouds that may host relatively larger accelerators. The top most layer is the Cloud, which has virtually unlimited powerful resources, but may be more distant from the user. 

Edge nodes are designed to be power efficient since they have a limited power budget. Hence, the accelerators on Edge nodes may be small and weak. To use accelerators on an Edge node efficiently, multiple virtual instances of the accelerator on the Edge node (similar to multiple VMs of the same processor) may be used to share the accelerator among multiple applications. We refer to the technology that facilitates the use of multiple virtual instances of the accelerator on a node as `accelerator virtualization'. 

Since there may be relatively fewer power restrictions on micro-clouds, they can host larger accelerators. Given the resource limitations on an Edge node, the accelerator virtualization technology can enrich computational capabilities available on an Edge node by providing access to accelerators located in a micro-cloud or even in the Cloud. These accelerators are remotely located and we therefore refer to this as `remote accelerator virtualization'~\cite{CCPE2017}. 

\vspace{2pt}
\begin{tcolorbox}[
    width=0.49\textwidth,
    enhanced,
    drop fuzzy shadow southwest,
    colframe=blue!50!black,
    colback=blue!2,]
\textcolor{black}{When the remote accelerator virtualization technique is leveraged, the CPU and accelerator components of the same application may be executed on different nodes. For instance, an application running on an Edge node may use powerful accelerators located in the micro-cloud or even in the Cloud instead of using weaker accelerators on the Edge node.} Both accelerator and remote accelerator virtualization optimize the use of accelerators in a typical resource-scarce and resource-constrained environment, such as the Edge.
\end{tcolorbox} 

Current Fog research usually focuses on generic processors and does not extend the use of accelerators to the Edge. The OpenFog reference architecture notes the benefit of accelerators for the Edge, but there is limited discussion on how accelerators can be made accessible on the Edge~\cite{openfog-1}. The focus of this article is to consider the research challenges and opportunities in making accelerators accessible at the Edge via virtualization for quickly processing user requests. 

\begin{figure*}[!ht]
  \centering
  \includegraphics[width=0.82\textwidth]{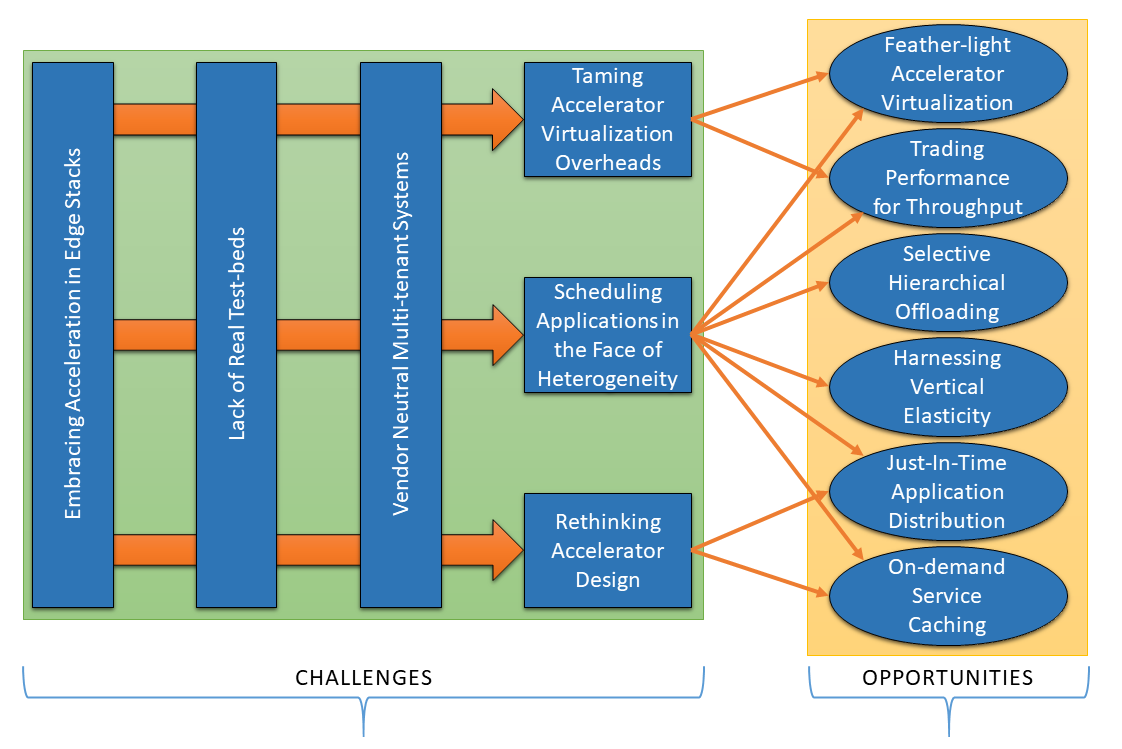}
  \caption{Challenges and the opportunities in making accelerators accessible in Fog computing.}
  \label{fig:challenges_and_opportunities}
\end{figure*}

\section*{Challenges}
\label{sec:challenges}
The main challenge in making accelerators accessible on the Edge is to holistically view the Fog system. Current research has developed independent subsystems within the Fog that are not interoperable and cannot be easily connected together. In the context of accelerators for the Edge and their use via virtualization, we anticipate that six challenges presented in this section will need to be addressed. Figure~\ref{fig:challenges_and_opportunities} highlights the challenges and how they map on to the opportunities that are considered in this article. The first three challenges are generic and influence the remaining three challenges. 

\subsection*{Challenge 1 - Embracing Accelerators in Edge Stacks}

Typically, the CPU is the primary consideration in the development of technology stacks. Support for accelerators in the stack is bolted on as an after thought, often resulting in a less adaptable core technology. The design of an end-to-end technology stack from the Cloud down to the Edge is flawed by the same approach. Therefore, the design approach needs to be changed by taking accelerators into account as primary processors or co-processors in the computing stack.

One potential avenue to explore in Edge stack design would be the inclusion of a decision layer and a scheduling layer at the bottom of the stack. The decision layer, operating at a macro-level, would be used to identify whether the service using an accelerator is available on a given Edge node. The scheduling layer, operating at a micro-level, would decide whether the accelerator could actually be made available for an incoming request based on system properties, such as resource availability. These layers should be designed considering the requirements of the broader Fog architecture in order to implement optimal decision-making algorithms at the decision and scheduling layers. %The challenge here is to implement a stack using a holistic approach. %This is further considered alongside the fourth challenge. 

\subsection*{Challenge 2 - Lack of Real Test-beds}
Research on Edge environments is rapidly evolving, but real practical physical test-beds are not yet available. We note that current research is fragmented in the following ways. Firstly, middleware and hardware solutions for potential Edge nodes are available, but end-to-end solutions are sparse at best. \textcolor{black}{Nonetheless as a starting point, technologies such as OpenStack Edge~\cite{openstack-edge}, OpenStack++~\cite{openstack++} and AWS Snowball~\cite{aws-snowball-edge} that support the use of the Edge are a starting point. These tools can be used to provide insight on the behaviour of workloads in the Fog architecture.} Secondly, simulation frameworks available for Edge environments are recycled in that they were once employed to study the Cloud and are now extended for the Edge. Given that there are no practical test-beds yet, the simulation models cannot be validated. Thirdly, experimental evaluation presented in the literature usually employs  bespoke and private test-beds that are crafted for specific applications. While this provides insight into how an application may perform, it is not fully representative of Edge infrastructure that is envisioned to be publicly accessible in the future. 

Consequently, it is difficult to understand from a research perspective the scale of applications or the resources really available in the Edge ecosystem although the business case for Fog computing is compelling and already well articulated. This translates into a lack of knowledge about the behaviour (communication between different levels at the Edge, between the Edge and the Cloud, and between the Edge and the users as well as computation that occurs at each layer) of real workloads. This obviously affects how a real application can be distributed across the Edge and the Cloud accelerators for maximizing computing performance but at the same time reducing the overall end-to-end energy consumed. Understanding these trade-offs will remain a challenge until end-to-end experimental test-beds can be publicly accessed. %It is challenging to understand the above trade-offs in face of fragmented research without any real test-bed.

\subsection*{Challenge 3 - Vendor Neutral Multi-tenant Systems}
Ideal service oriented systems, including Edge-based systems, are multi-tenant; that is, they host different services and serve multiple users. 
%This makes Edge-based systems inherently multi-tenant thus being able to host multiple services and serve multiple users. 
Conceptually, in the context of making accelerators accessible, virtualization at the Edge is a way to achieve multi-tenancy. However, this is a multi-faceted challenge: Firstly, there are no generic solutions to share accelerators, for example GPUs, among multiple users or applications. Current solutions available in the market are targeted for vendor specific GPUs. Vendor neutrality is a goal that remains to be achieved for GPU virtualization solutions and is essential for multi-tenancy on the Edge. Vendor neutrality may be achieved via API forwarding, which is a technique used to provide wrappers for functions which make use of the accelerator. These wrappers forward all function calls to the appropriate GPU. 

However, API forwarding raises a second challenge in that additional middleware is required on the Edge nodes. Because Edge nodes are likely to host small and low power accelerators, the more powerful accelerators will instead need to be 
located on micro-clouds. Although API forwarding would be necessary to reap the benefits of remote GPUs, the communication overheads would need to be minimized. Additionally, developing the technology for virtualizing small GPUs without significant resource footprint remains an open challenge.   

\subsection*{Challenge 4 - Scheduling Applications in the Face of Heterogeneity}
Edge environments are anticipated to be heterogeneous in at least two ways. Firstly, the Edge may comprise nodes ranging from resource constrained individual nodes to small clusters, such as micro-clouds. Secondly, multiple CPU and accelerator architectures may be used at the Edge. This however raises an important question - 'how can applications be distributed and scheduled by taking heterogeneity into account?'. This is challenging because there are multiple different combinations of layers and nodes in an Edge environment that could potentially be used by a scheduler to distribute an application. Scheduling becomes complex because weak and powerful accelerators co-exist at the Edge. In addition, the network conditions and the state of resources (who resides on the resources and for how long) will need to be monitored without consuming large amounts of resources at the Edge. 

API forwarding is one solution that will facilitate scheduling \textcolor{black}{of accelerators}. This is because the middleware can intercept all accelerator function calls made by an application and forward them to a suitable \textcolor{black}{remote} accelerator. Given that the middleware will be aware of the resource requirements, a monitoring entity that aggregates the requirements from all applications at geographically nearby Edge nodes may be used to create a more global view of the requirements and resources available. A scheduling plan could then be generated that would reduce response times and guarantee Quality-of-Service (QoS), so if an accelerator that hosts a distributed application cannot satisfy the QoS requirements, then 
\textcolor{black}{the component of the application that is to be hosted on an accelerator} may be migrated to a different accelerator. However, monitoring resources and migrating applications for scheduling applications are both open challenges.

\subsection*{Challenge 5 - Taming Accelerator Virtualization Overheads}
Virtualization inevitably introduces communication overheads, and due to the use of additional middleware, the same is true for accelerator virtualization. Although the communication overhead between CPUs and a GPU in the same node may be negligible because the PCI Express bus is used, this is not the case when remote accelerator virtualization is employed. 
%In the Edge, small accelerators may be located on Edge nodes and relatively larger accelerators may be available on Edge micro-clouds. 
The overheads will be compounded when a large accelerator required by an Edge node is physically located in the micro-cloud. The network bandwidth between the Edge node and the micro-cloud will significantly influence the overall performance of the virtualized accelerator. For example, a large number of data transfers between the CPU of an Edge node and a remote GPU in the micro-cloud could potentially negatively affect the overall performance of the application. The challenge in adopting accelerator virtualization on the Edge is to keep overheads to a minimum for meeting the QoS demands of an application. 

In supercomputing facilities and centralized Cloud data centers, the network overheads of virtualization are mitigated by the use of high-performance interconnects, but using these interconnects in a global Edge context is simply not feasible. Features such as Remote Direct Memory Access (RDMA), which are used in high-performance interconnects, could be adopted to reduce communication overheads between Edge nodes and micro-clouds~\cite{Middleware2015}. This would save CPU cycles and optimize end-to-end energy consumption. Nonetheless, efficiently incorporating RDMA in the Fog architecture may pose new challenges beyond the scope of this article. 

\subsection*{Challenge 6 - Rethinking Accelerator Design}
\label{memory_focus}

A key consideration for accelerator designers has been to improve bandwidth between computing cores of the accelerator and global memory, commonly referred to as data transfer bandwidth. Maximizing the data transfer bandwidth is essential for meeting performance requirements of High-Performance Computing (HPC) applications.
%that have a large number of data transfers between the CPU and the GPU. For example, an experiment we conducted using 4 million particles on the LAMMPS (Large-scale Atomic/Molecular Massively Parallel Simulator) benchmark~\cite{LAMMPS} spends nearly 30\% of the overall time on data transfers between the CPU and the GPU. 

However, HPC applications are not representative of the workloads that will utilize the Edge. The motivation for using the Edge is to facilitate computing closer to user devices and augment computing of user devices. Therefore, the class of applications best-suited for making use of the Edge will be user-oriented\textcolor{black}{, such as AR applications, for instance.}
%and may leverage machine learning algorithms. 
These applications are inherently different from HPC workloads \textcolor{black}{and therefore some accelerator characteristics that are necessary for HPC workloads, such as high data transfer bandwidth may not be so necessary in Fog computing due to the inherently different nature of the workloads executed.}

%in that there may be fewer data transfers between the CPU and GPU, making the data transfer bandwidth inconsequential. \textcolor{red}{*Comment 1* For example example, in an experiment we conducted, a distributed image classification application using Alexnet~\cite{alexnet} required only 1\% of the overall time for data transfers when classifying an image. The application workflow included training the model on the Cloud using a large GPU and sending a trained model to an Edge node. Users generated images and sent them to Edge nodes for classification instead of to the Cloud. }

\textcolor{black}{For example,} 
%The Edge node in the above example provided the acceleration required for classifying the image via an on-chip GPU integrated with the processor. These 
GPUs in Edge nodes are small and have low power consumption, \textcolor{black}{and, additionally, are not expected to accelerate HPC applications. Thus}
%but are less useful for accelerating machine learning workloads. 
it would seem that the primary design objective in this context should not be simply improving features, such as the data transfer bandwidth. Instead, designers should seek to surmount the challenge of miniaturising more powerful GPUs and making dedicated accelerators for workloads actually executed in this context.
%, such as Tensor Processing Units (TPUs) available on Edge nodes to operate within suitable power ranges. 
Looking forward, the challenge is to move away from the use of generic accelerators, instead designing Edge specific accelerators for targeted application domains.

\section*{Opportunities}
\label{sec:opportunities}
In light of the challenges, there are six avenues of opportunity for realising a complete end-to-end solution for using accelerators in Fog computing. Accelerator virtualization is an enabler for all opportunities and the solutions will need to take a holistic view of the Fog architecture to avoid fragmented research. 

\subsection*{Opportunity 1 - Feather-light Accelerator Virtualization}

Current solutions that offer accelerator virtualization for large-scale systems, such as Clusters or Clouds, assume the availability of significant compute resources for management activities that support virtualization. These management activities comprise a collection of tasks, such as monitoring the performance of the system to meet QoS demands, scheduling resources for an application that needs to be hosted, and scaling resources based on service demand. These tasks typically require significant amount of resources, we refer to as heavyweight, and are usually placed on an observer system. 

\vspace{2pt}
\begin{tcolorbox}[
    width=0.49\textwidth,
    enhanced,
    drop fuzzy shadow southwest,
    colframe=blue!50!black,
    colback=blue!2,]
While accelerator virtualization is demonstrated to work reasonably well on Clouds where there is an abundance of resources, it may not be pragmatic to make use of heavyweight management tasks on resource scarce Edge nodes. Therefore, a lightweight approach to accelerator management is required for the Edge.
\end{tcolorbox}

%In addition to lightweight management, the accelerator virtualization technology itself needs to be lightweight. 
However, since the Edge could be resource constrained, a lightweight approach is required. Not only should management be lightweight, but the accelerator virtualization technology also needs to be lightweight. There are two main techniques that are currently used for accelerator virtualization, namely using virtual drivers and API forwarding. Both techniques require full knowledge of the accelerator drivers. The first technique is completely platform dependant and is less flexible in dynamically accommodating more clients. The second technique usually offers a more comprehensive solution, but is complex and does not easily work for proprietary accelerator drivers. This hinders portability across different accelerator platforms and consequently their widespread adoption.  
\textcolor{black}{In light of the above drawbacks and given that resources required for meaningfully virtualizing accelerators have not been a major concern on Clouds and clusters given the availability of significant resources, current accelerator virtualization solutions may not be fit for direct use on the Edge. Edge nodes do not have the same computational capabilities as centralized resources and therefore development of a lightweight accelerator virtualization technology is essential.}
We identify opportunities in adopting a two-fold fundamentally different design approach for developing accelerator virtualization for the Edge. Firstly, by developing accelerator virtualization mechanisms that focus on flexibility so that clients can be added or removed to cope with dynamic workloads as seen in Edge settings. Secondly, by developing platform and driver agnostic solutions to integrate a wide variety of accelerators, representative of the Edge ecosystem, for achieving vendor neutrality.
\textcolor{black}{  
Adopting the above design approach, will help materialize lightweight accelerator virtualization solutions and make them more suitable for the Edge.}

\subsection*{Opportunity 2 - Trading Performance for Throughput}
%Even a lightweight accelerator virtualization solution will have overheads.
In practice, virtualization could translate to reduced application performance because an accelerator is shared among different applications. Ideal acceleration is achieved when an application can exclusively use the accelerator on an Edge node or on a micro-cloud. But this may lead 
to reduced overall throughput.

\vspace{2pt}
\begin{tcolorbox}[
    width=0.49\textwidth,
    enhanced,
    drop fuzzy shadow southwest,
    colframe=blue!50!black,
    colback=blue!2,]
Exclusively using an accelerator for an application does not necessarily mean that an ideal overall throughput is achieved. This is because applications might use accelerators for very short time intervals and may consume a very small fragment of the accelerator resources.
\end{tcolorbox} 

For example, \textcolor{black}{image processing algorithms on moderate and large GPUs, such as used for the AR use-case, may not utilize the entire GPU memory.} Moreover, when applications need to be accelerated, they typically wait until the preceding application has finished using the GPU. There is opportunity here for developing frameworks to support concurrent execution of applications via multiple virtual GPUs to improve accelerator utilization and service multiple applications on the accelerator at the same time. This results in a higher overall throughput without noticeably degrading the performance of individual applications~\cite{gpucoscheduling}. Additional research is required to develop and implement co-scheduling strategies on Edge nodes and to understand the impact of considering the overall throughput as against individual performance of applications for meeting QoS objectives.

\subsection*{Opportunity 3 - Matchmaking in the Fog Hierarchy: Selective Hierarchical Offloading}
The Fog hierarchy is described in terms of layers with different computing capabilities in Figure~\ref{fig:fog}. 
%At the bottom is the user device layer, followed by the Edge layer\st{s} comprising nodes that may have a weak accelerator and micro-clouds with more computing power. The Cloud layer has virtually unlimited computing power.
Contrary to servers in a Cloud data center, power is an essential design criteria for Edge nodes. Unlike data centers that host a large amount of resources, only limited computing resources can be incorporated in an Edge node. 

\vspace{2pt}
\begin{tcolorbox}[
    width=0.49\textwidth,
    enhanced,
    drop fuzzy shadow southwest,
    colframe=blue!50!black,
    colback=blue!2,]
While in principle it would be advantageous to have large amounts of resources to process data closer to the source at the Edge, it is practically impossible due to power limitations at the Edge. For example, the Thermal Design Power (TDP) of a home router will be at least a magnitude lower than that of a large accelerator. This is a fundamental problem that cannot be easily addressed.
\end{tcolorbox} 

Since Edge nodes are power limited, an application will need to be carefully distributed across different layers for improving performance. This requires the selection of accelerators in the right layer for offloading an application, which if not carefully selected would be detrimental to performance of the application. This naturally provides two avenues of opportunity in scheduling. \textcolor{black}{The first avenue, related to the initial deployment of an application, is referred to as selective hierarchical offloading. This initial deployment can later be enhanced by the second avenue, referred to as vertical elasticity, 
which dynamically accounts for redeploying the accelerated component of an application if required (this is considered further in Opportunity 4)}. 

We identify that opportunities for selective hierarchical offloading in the Fog may be based on both functional and non-functional properties. Functional properties may include \textcolor{black}{the computational intensity of the application, which indirectly translates to} execution time. If an application is expected to take more than an acceptable target time on the accelerator in an Edge node for servicing user requests, then remote accelerators may be used from the micro-cloud. Non-functional properties may include the consideration of the class of applications. For example, the network overhead may be less significant for compute bounded applications and therefore remote accelerators may be beneficial. On the other hand, bandwidth bounded applications may be affected by network overheads and offloading onto accelerators in multiple layers may not be beneficial. This will need to be underpinned by novel scheduling strategies that select accelerators in the right Fog layers to meet service objectives. 

\subsection*{Opportunity 4 - Matchmaking in the Fog Hierarchy: Harnessing Vertical Elasticity}

\vspace{2pt}
\begin{tcolorbox}[
    width=0.49\textwidth,
    enhanced,
    drop fuzzy shadow southwest,
    colframe=blue!50!black,
    colback=blue!2,]
In contrast to horizontal scaling, which is adding resources of the same type used in Cloud data centers, we recommend vertical scaling for using the Edge. \textcolor{black}{We define vertical elasticity in the context of this article as automatically scaling a workload across the Cloud-Edge continuum, such that an Edge node may remotely utilize accelerators in micro-clouds or even the Cloud for satisfying QoS requirements}. For example, if an application cannot be serviced by a GPU on an Edge node, then a larger GPU or a collection of GPUs from an Edge micro-cloud may be accessed to service the request.  
\end{tcolorbox} 

We anticipate opportunities in the development of at least two different approaches for achieving vertical elasticity. In the first approach, when an Edge node receives a request that it cannot satisfy with its own accelerator, the Edge node may simply forward the request on to a micro-cloud. \textcolor{black}{We assume that micro-clouds are closer to user devices than the Cloud. In this manner, we may still achieve low latency.} In a second approach, the Edge node may use remote accelerator virtualization, for example, to use the accelerators on the micro-cloud to service the request without forwarding the incoming request to the micro-cloud. In the former approach, the Edge node is simply a proxy and the network traffic is sent as far as the nodes that can satisfy the request. \textcolor{black}{Therefore, the entire application is moved from an Edge node to the micro-cloud}. In the latter, however, the Edge node services the request by making use of accelerators that are located elsewhere in the Edge ecosystem. \textcolor{black}{Hence the CPU component of the application is still executed on the Edge node but computational intensive components that require the accelerator are offloaded to a remote accelerator, potentially in the Edge micro-cloud.}
These approaches will need to be further investigated for fully understanding their benefits for vertical elasticity. 

\subsection*{Opportunity 5 - Just-In-Time Application Distribution}
The execution environments that support distributed applications usually make decisions during compilation time on distributing the application via partitioning across multiple nodes or heterogeneous hardware and where each partition must be executed. 

\vspace{2pt}
\begin{tcolorbox}[
    width=0.49\textwidth,
    enhanced,
    drop fuzzy shadow southwest,
    colframe=blue!50!black,
    colback=blue!2,]
Support for distributing an application across heterogeneous resources is available via the programming model and programming language. The onus of determining an (or the most) efficient distribution currently rests with the programmer. Just-In-Time distribution may overcome this burden.
\end{tcolorbox} 

Runtime solutions in this space may cope with variable sizes of input data for each partition, but research in partitioning an application based on the availability of different hardware or even environment conditions, such as network traffic, is not mature. Runtime solutions may either search the problem space exhaustively to determine an efficient distribution plan or may predict performance of different combinations of partitions using a mathematical model. This is cumbersome and may not be suitable for resource constrained and dynamic Edge environments.

There are numerous opportunities in the Edge context to research runtime distribution of applications, we refer to as `Just-in-Time' (JIT) distribution. These include developing mechanisms to offload the parallel component of an application to one or more virtual and/or remote accelerators. The JIT distribution could be based on reducing total execution time of the application, finding optimal geographical locations for placing partitions and availability of resources for fulfilling QoS of the application, minimizing response times, and improving the overall energy efficiency and performance. These mechanisms may be used on a single node that contains an accelerator or across multiple nodes to access remote virtual accelerators.
However, a key question that will still need to be addressed is how to host individual partitions on accelerators with very little overheads in the Edge ecosystem.
\textcolor{black}{In the context of CPU only applications, container technology, such as Docker is widely used to package and deploy partitions across resources. Containers are yet to become fully interoperable across heterogeneous resources. Although containers have lower overheads than VMs, in practice they require time in the order of a few seconds to start up. Accelerator supporting containers have started to emerge, such as GPU-Enabled Docker containers, but are still in their infancy~\cite{gpucontainers}. Container overheads will need to be reduced or a radically different interoperable technology similar to containers will need to be developed. This will facilitate JIT distribution of applications.}
%In addition, containers are used to deploy user written functions, which are typically short running, on a suitable Cloud resource without requiring input from the user on which resource needs to be used. This is referred to as} `Serverless' computing~\cite{serverless-1}. 
%In this computing model, an execution environment for executing smaller units of a program (such as functions) may be one solution to surmount the problem~\cite{serverless-1}. 
%\textcolor{blue}{In this way it should be defined, for instance, how to incorporate in serverless computing such mechanisms in order to optimize the execution when distributing applications through a number of heterogeneous resources. One possible way could be doing an implicit profiling to characterize the applications. With this characterization it would then be possible to properly distribute the applications to obtain optimal performance.}

\subsection*{Opportunity 6 - On-demand Service Caching}
%Data caching is a well known technique in Cloud computing whereby data in high demand is redundantly stored in a location geographically closer to where it may be used (instead of only using a remote data center) for reducing latencies. Similarly, 
We identify service caching as an opportunity to reduce communication latencies and optimally make use of accelerator resources at the Edge. \textcolor{black}{That is,} services in high demand making use of accelerators
%, such as machine learning applications, 
may be cached on Edge nodes. However, not all services can be cached given that \textcolor{black}{accelerator} resources available on an Edge node \textcolor{black}{are} limited. Hence, the obvious question needs to be answered assuming \textcolor{black}{accelerator} multi-tenancy - which services should be cached on an Edge node?

%Accelerator virtualization allows for a  
%A single Edge node may not be able to cache all available services given the resource limitation. However
%more efficient
%ly 
\textcolor{black}{Accelerator virtualization decouples the CPU and accelerator components of an application thus allowing for a more efficient service caching}
in the following two ways. Firstly, by developing mechanisms to \textcolor{black}{move computation to} remote accelerators on micro-clouds \textcolor{black}{so that they} host services that cannot be cached on the Edge node. Secondly, by developing methods to migrate services \textcolor{black}{that require acceleration} between Edge nodes and micro-clouds based on demand. In both cases, only the accelerator component of 
the application is moved, whereas the CPU component remains on the Edge node. 
For instance, less popular services may be cached on micro-clouds whereas more popular services may reside on Edge nodes. If user demand for a service increases, then it may be migrated to the Edge node and swapped for a less popular service. Service caching in Fog computing is currently untapped into and remains an open avenue to be explored.

\vspace{2pt}
\begin{tcolorbox}[
    width=0.49\textwidth,
    enhanced,
    drop fuzzy shadow southwest,
    colframe=blue!50!black,
    colback=blue!2,]
Accelerator virtualization facilitates service caching on Edge nodes. \textcolor{black}{Instead of caching the entire application, only the accelerator component of the application is cached. This is possible because the CPU and accelerator components of an application are decoupled. Therefore, the accelerator component can be moved to any remote accelerator while the CPU component resides on the Edge node.} 
\end{tcolorbox} 

\section*{Summary}
\label{sec:conclusions}
%Bringing computing closer to the edge of the network is appealing for achieving a sustainable Internet. Using the Cloud along with the Edge, referred to as Fog computing, is a research area still in its infancy. In this article, we have discussed the potential of incorporating hardware accelerators at the Edge for improving the performance of services that will be hosted on the Edge. The challenges and opportunities arising from this relatively new direction of research are highlighted. Maintaining an end-to-end perspective of Fog architecture, from design to delivery, is essential for conducting useful research in this area. 

\textit{How can enhanced analytics be possible at the Edge in Fog computing?} Application specific hardware accelerators designed for Edge nodes are key to achieving this and accelerator virtualization is an enabling technology to facilitate a service oriented Edge. Looking forward, maintaining an end-to-end perspective of the Fog architecture, from design to delivery, is essential for conducting useful research in this area.

\bibliographystyle{IEEEtran}
\bibliography{references}

\end{multicols}

%\newpage
%\section*{Author Biography}
%\vspace{-18pt}
%\input{biography}

\end{document}